# Stabilizing ultrathin Silver (Ag) films on different substrates

**Allamula Ashok[1], Pradeep Kumar Rana[1], Daljin Jacob[1], Peela Lasya,[1] P Muhammed Razi[1] and Satyesh Kumar Yadav[1,2]***

1 Materials Design Group, Department of Metallurgical and Materials Engineering, Indian Institute of Technology (IIT) Madras, Chennai, India.

2 Center for Atomistic Modelling and Materials Design, Indian Institute of Technology (IIT) Madras, Chennai 600036, India.

ashok.allamula@gmail.com, satyesh@iitm.ac.in[*]

**Abstract**

This paper reports an effective method of stabilizing ultrathin Silver (Ag) films on substrates using a filler metal (Zn). Ag films with a thickness < 15 nm were deposited by DC magnetron sputtering above a Zn filler metal on glass, quartz, silicon and PET (polyethylene terephthalate) substrates. Zinc is expected to partially or fully fill the roughness associated with the substrates. The Zn filler material and ultrathin Ag film form a 3-D augmented atomically chemically graded interface. 3-D interfaces have smoothly varying chemistry. The ability of Zn to partially or fully fill the substrate roughness improves the adhesion of Zn along with the Ag to the substrate. Also, Zn acts as a barrier layer against the diffusion of Ag into the substrate. This technique leads to ultrathin Ag films with low sheet resistance (~ 3 Ω/Sq.), low mean absolute surface roughness (~1 nm), good optical transparency (~ 65 %), better stability and compatibility with the environment. The results indicate significant potential for applying stable ultrathin Ag film/electrode as a practical and economically feasible design solution for optoelectronic (transparent and conductive electrodes for solar cells and LEDs) and plasmonic devices. This film shows good conductivity, transparency, stability, and flexibility.

**Keywords:** ultrathin Ag films, magnetron sputtering, sheet resistance, surface roughness, filler metal, Silver, Zinc.

* Corresponding author: Dr. Satyesh Kumar Yadav

**Introduction**

Flexible electronics have a wide range of applications in flexible electrodes (top and bottom electrodes for solar cells, LEDs), foldable displays, flexible circuits and wearable sensors for healthcare [1], [2]. Modern flexible electronics need materials with good optoelectrical properties and mechanical flexibility [3], [4]. High electrical conductivity and optical transmittance are the main factors that determine the quality and performance of the electrodes [4], [5]. Indium Tin Oxide (ITO) is widely used as a transparent electrode in solar cells and Light Emitting Diodes (LEDs) possessing good broadband transmittance (~80%) and low sheet resistance (~10 Ω sq$^{-1}$) [3]. However, growing expense and the intrinsically brittle nature of ITO electrodes which results in these materials are undesirable for future flexible electronics [6]. Several attempts are being made to replace them.

Due to the shortcomings of ITO, various research efforts are being made to find an alternative transparent conductive electrode. These can be fabricated using a cheaper and more reliable process having reasonable optoelectrical characteristics and robust structural durability for flexible and stretchable applications [3]. Graphene, carbon nanotubes (CNTs), conductive polymers, metal nanowires, metal meshes, and ultrathin metal films have been used to make ITO-free optoelectronic devices resulting in higher performance. Challenges remain due to non-uniform optoelectronic properties of CNTs due to their random distribution, lower conductivity of polymeric Polythiophene 3, 4-ethylene dioxythiophene: Polystyrene sulfonate (PEDOT: PSS) based films due to poor chemical stability, and rough surface of metal nanowires on CNTs (Ag nanowire/CNT; Ag nanowire/PEDOT) based films, makes these materials incompatible as transparent electrodes in optoelectrical devices [5].

Ultrathin metal films exhibit good optoelectrical properties, flexible, lightweight, stretchable and can withstand bending cycles without any optoelectronic property variations. Therefore, ultrathin metal films can be a viable candidate for optoelectrical devices [5]. The deposition of a metal layer follows the Volmer–Weber nucleation mode due to its poor adhesion with the substrate (mismatched surface energies). This results in a non-continuous film with discrete islands and random crevices that increases surface roughness. Reduction in optical transmittance and electrical conductivity are consequences of surface irregularities [7], [8]. The competition between adatom-adatom interaction and adatom-surface interactions controls the growth of ultrathin films. The thickness of the silver film plays a crucial role in optoelectrical properties. If the thickness is too small (<10 nm), outcomes multiple discontinued nano-islands rather than forming a flat thin film [4].

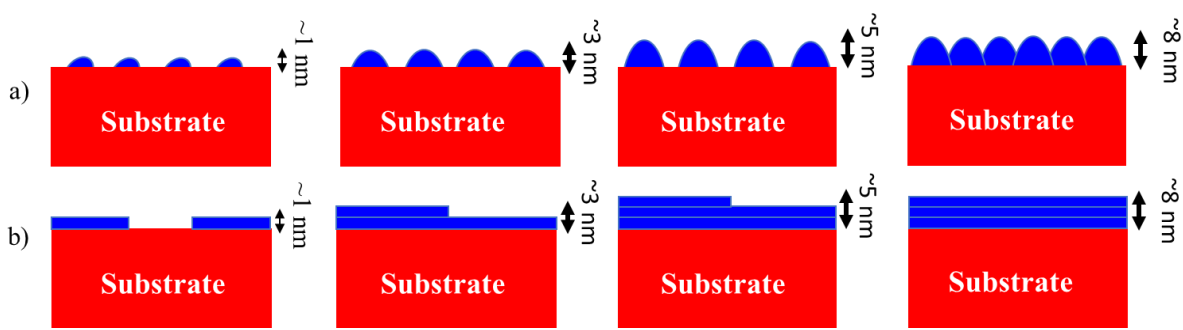

**Fig. 1.** Typical nucleation of a thin metal film in initial stages on the substrate (a) Volmer-weber mechanism results in the formation of nano-island below critical thickness (b) Frank –van der Merwe, growth takes place monolayer by monolayer.

Silver thin films have a broad spectrum of applications due to their better performance than other metal films. Silver (Ag) is known for its excellent electrical and optical properties; lowest electrical resistivity, highest optical reflectance, and the lowest emissivity among metals [9]. Ag thin films are widely used in optical devices [10][11], solar devices [12], sensors [13], electronics [14], and plasmonic devices [15]. Depositing an ultrathin film of an Ag (noble metal) on a ceramic substrate has been in consideration for a few years. However, the most common issue being faced as such is the formation of nano-islands of metal instead of continuous layers. Considering the interfacial energy mismatch and inbuilt roughness can help us solve this issue.

Studies have reported that some approaches are followed to make ultrathin Ag films. Reportedly, they are alloying and or seed layer techniques [5]. These techniques include the deposition of very thin seed layer of; 0.1 nm thick Niobium (Nb) [16][1], 1-2 nm thick Germanium [17] [2], 5-6 nm of Germanium and Chromium [18] [3], Nickel [19], Molybdenum trioxide (MoO3) [20] [5], 0.8 nm thick PEDOT-PSS [21] [6], 1 nm thick Copper, Silicon, Titanium [7], [8] [7] between the Silver and substrate. Co-deposition of; up to 10 at% of Aluminium [22]–[24], Copper, Nickel, Titanium, Chromium [4], [5], [25] with Silver. Suppression of the island growth and favoring the layer-by-layer growth resulting in ultrathin, continuous, smooth films with good optoelectrical properties.

The seed layer and co-deposition techniques help to reduce the surface roughness of Ag films by acting as better nucleation sites for Ag film. Ultrathin Ag films deposited using Germanium, Aluminium, Copper as seed/co-deposition suffers from lower conductivity, diffusion of such layers into substrate and noble metal films [26]. Si/SiO2 is the commonly used substrate in most the electronic and optoelectronic applications. In these cases, the seed or co-deposition of Al, Cu reduces the substrate itself. These reports studied making transparent and conductive Ag (noble metal) films. In contrast, the underlying mechanism to stabilize the ultrathin metal (Ag) films still needs to be well established and yet to be investigated.

Therefore, a method/technique (modifier material) is needed to address the shortcomings discussed above and to make stable ultrathin Ag films that are viable for optoelectronic devices. <1 paragraph on DFT. DFT results suggests that Zinc has that unique ability to absolutely get into the nano crevices more efficiently and fill them up. The enthalpy of formation of neutral clusters is very low for the clusters of zinc with n atoms compared to that of other element clusters (Gold, Copper, Manganese) **Fig. 3.** with the same number of atoms.> In this study, a new filler metal (Zn) is used, which is nowhere studied and highlighted the reason why this filler material helps for the formation of a 3-D augmented interface and the underlying mechanism behind this is investigated. We observed/found that The Zn-assisted Ag film showed very good continuity (Fig. 3c) (low sheet resistance). This is due to the Zn filler metal; substrate and ultrathin Ag form a 3-D augmented atomically graded interface. These 3D interfaces improve adhesion between ultrathin Ag film, Zn, and substrate. The ability of Zn to partially or fully fill the roughness associated with the substrate improves the adhesion of Zn along with the ultrathin Ag film to the substrate. In this way, in ultrathin Ag stabilized device applications, the Zn acts as a barrier layer against the diffusion of ultrathin Ag into the substrate. Zn filler metal layer assists in reducing the roughness of the substrate by filling up the nano corrugations (nano crevices) in mentioned substrates. Wherein the ultra-thin metal film is impervious to the other layers on top or if exposed to the atmosphere, it is impervious to the atmospheric gases so that it does not react with the Zn filler metal. Nano island formation is due to an interfacial energy mismatch between the substrate material and the deposited metal. Instead, Zn filler material is used to fill up the nano-crevices or roughness in a material to make it flat and reduce the interfacial energy mismatch. Any metal can now be deposited on this flat substrate; a continuous layer can be formed even with minimal thickness. Thus, the objective of the present work is 1. To synthesize ultrathin metal films of silver with better electrical conductivity and optical transmittance. 2. To improve the stability of the films up to 300 Oc.

**Experimental Methods**

Glass, Quartz, Silicon substrates were used in this experiment. Prior to deposition, the substrates were cleaned using acetone and ethanol, followed by rinsing with distilled water and drying with 99.999% pure $N_2$ gas. The deposition processes were performed in ATS Hind High Vacuum 500 DC magnetron reactive sputtering system. Before deposition, the sputtering chamber was evacuated to $5\times10^{-6}$ mbar to ascertain the purity of the thin films. Pre sputtering for 10 min was carried out at $8.5\times10^{-3}$ mbar and 30 W DC power to ensure the target surface was free from contamination. Ag was sputtered from a 99.99 % pure target, with a 2 in. diameter and thickness of 0.25 in. Argon (99.999% purity) was used as the carrier gas for sputtering. Three deposition process parameters: sputtering power, working pressure, and deposition rate were optimized to deposit ultrathin Zn assisted Ag films and the deposition process parameters were mentioned in **Table 1**.

**Table 1:** The deposition process parameters of Ag and Zn target materials used in this study.

| Process parameter | Target Material (Ag) | Target Material (Zn) |
|---|---|---|
| Sputtering power (W) | 5 | 10 |
| Working pressure (mbar) or Flow rate (sccm) | 5×10$^{-3}$ (23) | 8.5×10$^{-3}$ (31) |
| Deposition rate (nm/min) | 1.5 | 2 |

The deposition was done at three different conditions 1. as deposition of Ag 2. as deposition of Zn and Ag 3. preheating of substrates at 100 °C for 30 min followed by Zn deposition and holding for 1 hour at 100 °C followed by cooling to room temperature (35 °C) then, Ag deposition and annealing at 100 °C inside the chamber for 1 hour. Hereafter, these deposition conditions will be referred to as condition 1, condition 2 and condition 3 respectively.

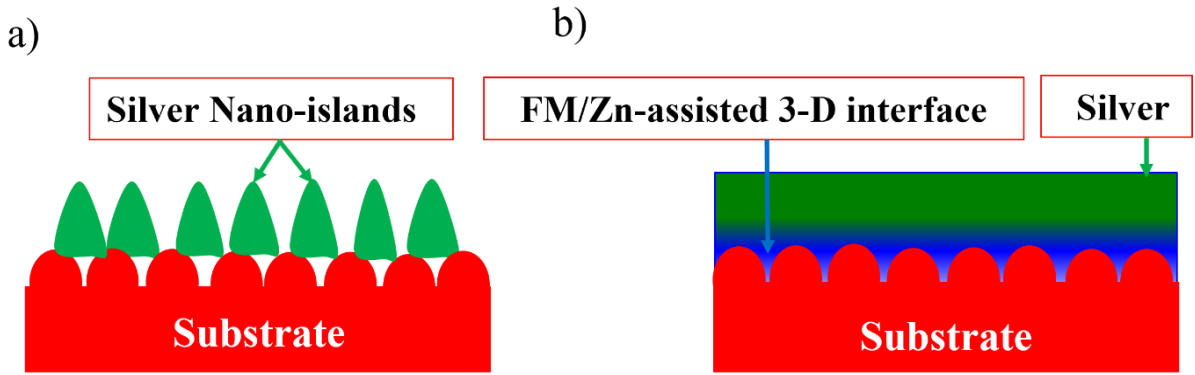

**Fig. 2.** Typical growth of ultrathin silver film (< 12 nm) (a) On ceramic substrate in the form of nano-islands (b) 3-D interface formation by filling of nano-roughness with filler material (FM)/Zn and smooth Ag ultrathin layer.

The surface roughness ($R_a$) of deposited Zn assisted Ag films was measured by an atomic force microscope (AFM) (BRUKER, Dimension icon) with the scan area of 3 μm × 3 μm and 10 μm × 10 μm for all the samples and the film thickness (t) is obtained by using an optical profilometer. The surface morphology of ultrathin Ag films was observed by scanning electron microscope (SEM) (Apreo S). The sheet resistance ($R_s$) is measured using a four-point probe (JANDAL, RM3000) with appropriate current supply and voltage measuring circuitry. The transmittance data was obtained by

UV-VIS spectroscopy (Perkin Emler LAMDA 950). Multiple readings were considered in each case, and the average and standard deviation were calculated.

## 3. Results and Discussion

### 3.1 Formation energy calculations

The nano-size effect plays a pivotal role in the interface stabilized phases [27]. With the size of the particles less than 10 nm, the dispersion of the nano-size increases and segregates at the gaps in the interface as amorphous-like clusters. These clusters collide with each other during diffusion at higher temperatures forming a layer at the interface and therefore affecting the thin film formation on top of it. To study this gap-filling ability of the Zn clusters, it is important to study the Zn clusters to comment on their stability.

Clusters of metals can take diverse shapes with some of the shapes energetically being more favorable in comparison to others [28]. This work reports, a comparative analysis of the formation of various clusters. The metal clusters of silver, chromium, Copper, Magnesium and certain transition element clusters of Niobium, Nickel, Zinc, and Manganese have been taken into consideration to ensure that all possible isomers have been explored exhaustively.

Formation energy per atom for n-atom cluster $M_n$ is given by:

$$\left(\frac{FE}{atom}\right) = \left[\frac{E(M_n) - n * E(M_1)}{n}\right]$$

Metal clusters have been studied extensively using DFT in the past two decades [29], [30][31][32] due to numerous applications in many fields including nanoelectronics. We generated the initial structures of these clusters using the VESTA software and studied all possible isomers of the cluster for n= 2-7 cluster sizes.

All calculations performed are optimized using Vienna Ab initio Simulation Package (VASP) [33] by the generalized gradient approximation (GGA) using the Perdew-Burke-Ernzerhof (GGA-PBE) exchange-correlation functional [33] and Projector Augmented Wave (PAW) datasets. All the vasp pseudopotentials used were basic ones without involving any p- or s- shell semi core electrons. For all the initial geometries, a 15Å length cubic unit cell was taken with clusters at the centre of the unit cell. Since a large unit cell was used, it was sufficient to use gamma (Γ)-point approximation for Brillouin zone sampling. SCF calculations are performed with a convergence criterion of $10^{-5}$ a.u. and the Conjugate gradient method (IBRION = 2) with a convergence criterion of $10^{-3}$ a.u. is used to perform all geometrical optimizations. Since the volume of the unit cell is kept fixed (ISIF = 2), ENCUT is taken as same as that of ENMAX from the POTCAR file with PREC = Accurate for all cluster calculations. Since the work focused only on relaxations in metals, Methfessel-Paxton smearing was used with ISMEAR = 1 as it gives the best results for metals.

The following Table 2 demonstrates the number of clusters of silver along with the 2D-symmetricity for these clusters. It is noted that the symmetricity for most of the metal is 2D for n= 2-11

**Table 2:** The symmetricity of the clusters used in this study.

| Number of clusters | Symmetricity |
|---|---|
| 2 | Linear |
| 3 | $C_{2v}$ |
| 4 | $D_{2h}$ |
| 5 | $C_{2v}$ |
| 6 | $D_{3h}$ |
| 7 | $C_S$ |

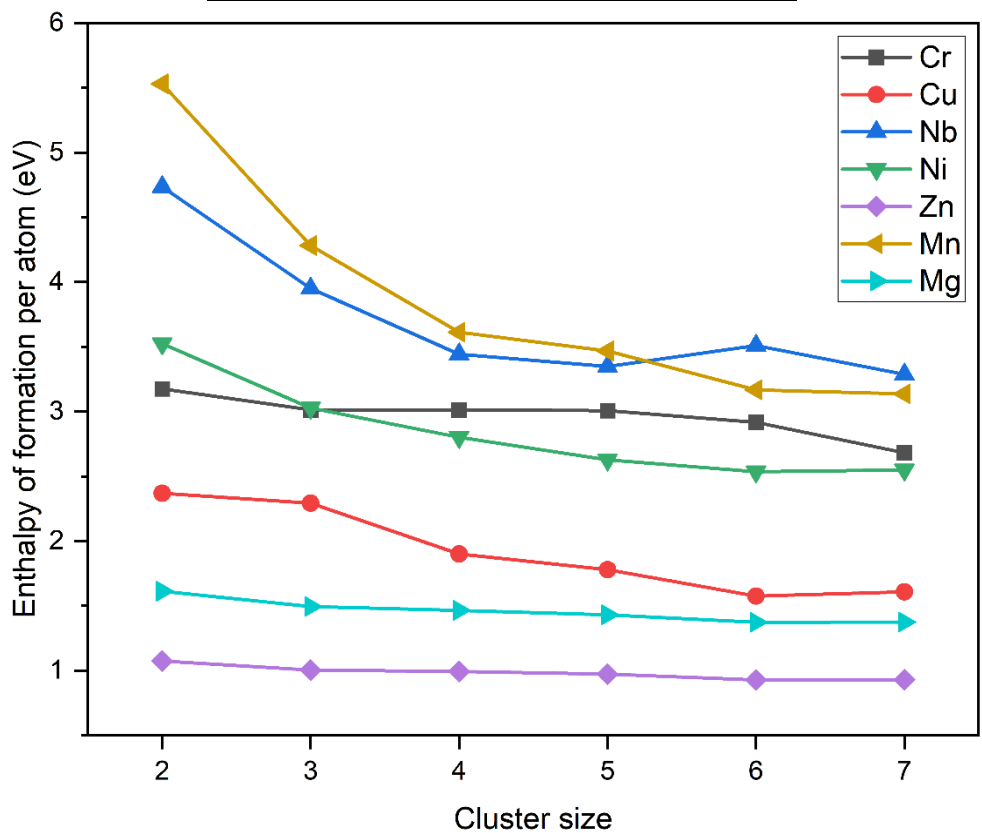

**Fig. 3.** Enthalpy of formation of neutral clusters as a function of cluster size.

The different alternatives available as an intermediate layer between the ceramic substrate and Ag for stabilizing ultrathin Ag films are Copper, Iron, Cobalt, Nickel, Chromium, Manganese, Zinc, Aluminium, Magnesium, Titanium, Niobium and Calcium. Out of all the available choices, the affinity of Titanium, Aluminium, Magnesium, and Calcium towards oxygen is higher than that of SiO2 (the most widely used dielectric). When these materials are used as filler material there is a higher possibility that they can reduce the underlying silicon oxide substrate to form their respective oxides. The elements like Copper, Iron, Cobalt, Nickel, Chromium, and Manganese have lower oxygen affinity than Silica and hence they won't reduce the underlying Silica. But the size of their neutral nanoclusters is larger compared to that of Zinc. Zinc has that unique ability to absolutely get into the nano crevices associated with silica substrates more efficiently and fill them up. The enthalpy of the formation of neutral clusters is very low for the clusters of zinc with n atoms compared to that of other element clusters (Cu, Mn, Mg, Ni, Nb, Cr) **Fig. 3.** with the same number of atoms. The melting point of Zinc is also the lowest among the candidate metals at around 420°C. Therefore, Zn is considered to be a filler material in this study to stabilize the ultrathin Ag films.

### 3.2 Characterization of thin films

### 3.2.1 Surface morphology

The surface morphology of Ag ultrathin films was observed by SEM. **Fig. 3**. Shows a) discontinuities on the surface of 6 nm Ag film due to the interfacial energy mismatch between Ag and substrate that leads to island growth of Ag film on top of the quartz substrate (open circuit) <cite> b) continuity on the surface of 6 nm Ag film this id due to 3D assisted growth of the film ($R_S$ of 60 $\Omega$ /sq.) c) The surface clearly shows the more continuity on the surface of Ag film (this is due to the prolonged annealing of the film) (Rs of 12 Ω /Sq).

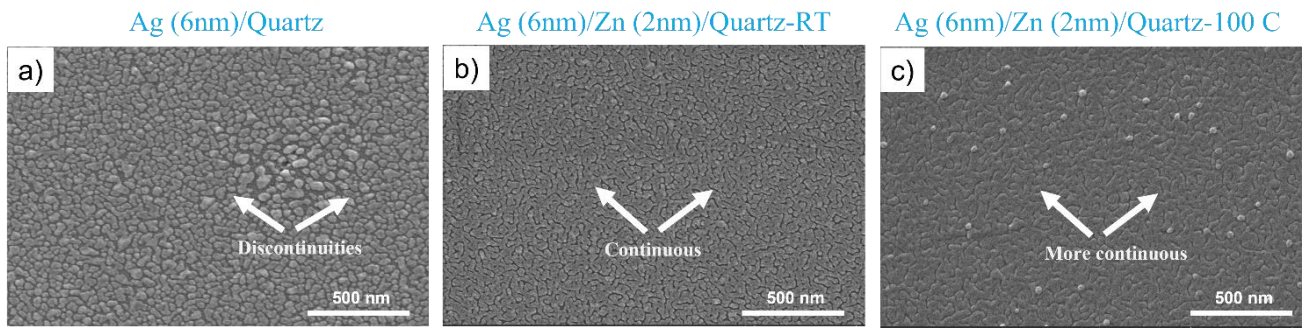

**Fig. 4.** SEM images of Ag thin films deposited on a quartz substrate (a) as deposited Ag (6 nm) film ;(b) as deposited Zn (2 nm) assisted Ag (6 nm) film ;(c) Zn assisted Ag film (preheating of substrates at 100 °C for 30 min followed by Zn (2 nm) deposition and holding for 1 hour at 100 °C followed by cooling to room temperature (35 °C) then, Ag (6 nm) deposition and annealing at 100 °C for 1 hour) or Zn (2 nm) assisted Ag (6 nm) film deposited as condition 3.

The Zn assisted Ag film deposited at condition 3 showed good continuity (Fig. 3c) (low sheet resistance). This is due to the Zn filler metal; substrate and ultrathin Ag form a 3-D augmented atomically graded interface. The same condition 3 was used to deposit and study the optoelectronic properties of ultrathin Ag films.

**3.2.2 Surface roughness**

The surface roughness of Zn filer metal 3D assisted Ag ultrathin films was measured by AFM. Fig. 2. a) and b) show two- and three-dimensional AFM images of stable Ag films on glass, quartz and silicon substrates at condition 3 (with preheating of substrates at 100 °C for 30 min followed by Zn 2 nm deposition and holding for 1 hour at 100 °C followed by cooling to room temperature (35 °C) then, Ag

8 nm deposition and annealing at 100 °C for 1 hour). The average surface roughness of these films was 2.57 nm, 2.09 nm and 2.02 nm respectively.

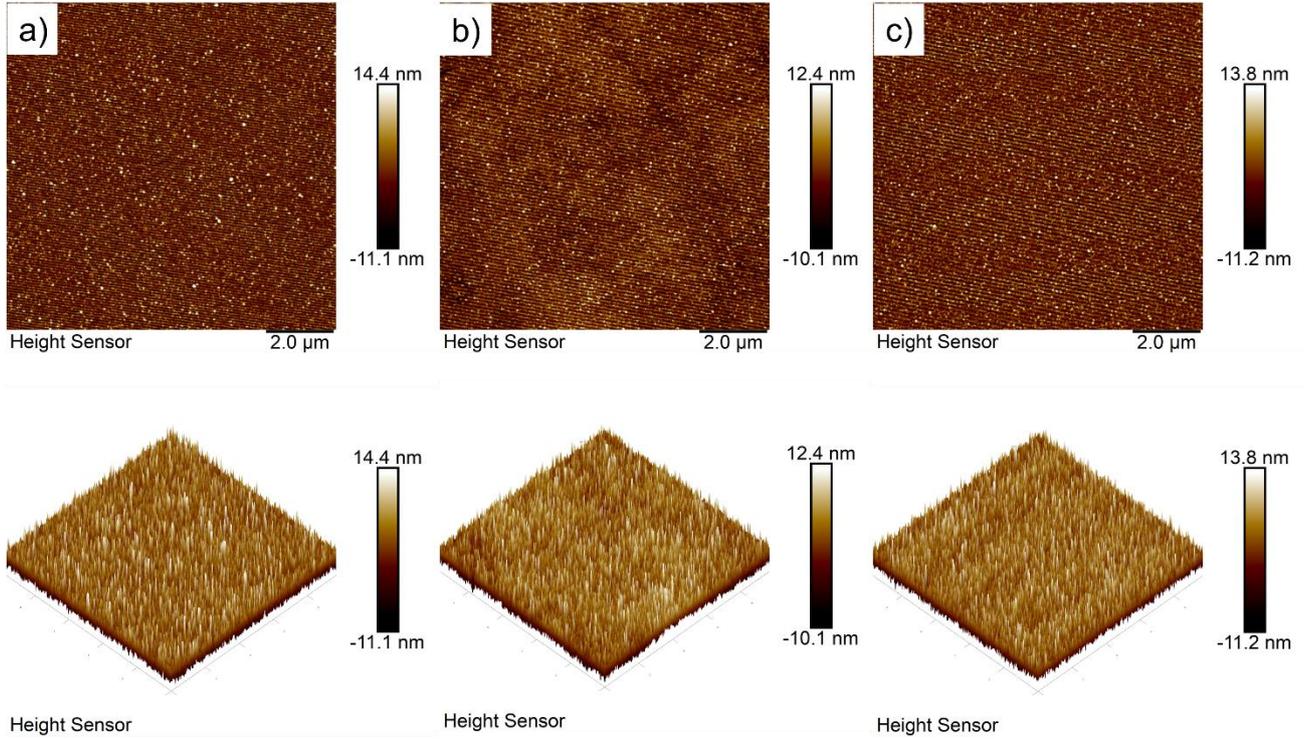

**Fig. 5.** AFM 2D and corresponding 3D morphology of Zn (2 nm) 3D assisted Ag (6 nm) films deposited on a) glass, b) quartz c) silicon with a scan size of 10 um * 10 um.

### 3.2.3 Optoelectronic properties

Ultrathin Zn 3D assisted Ag films were deposited on glass, quartz and PET substrates to obtain a transparent and conductive film. The sheet resistance and optical transmittance of the Ag films were measured and shown in Table **2**. The combined Haacke figure of merit (FOM) [34] is defined as

$$FoM = \frac{T^{10}}{Rs} \ldots \ldots (1)$$

where T is the transmittance, measured at 550 nm, and $R_S$ is the sheet resistance (Ω /sq.). The values are shown in **Table 2.** Increasing Ag thickness increases the sheet resistance but also reduces the transmittance due to the bulk nature of Ag which leads to increased scattering. The FoM is higher for the Ag (8 nm)/Zn (2 nm) films deposited on quartz.

**Table 2:** Haacke FoM of Zn 3D assisted Ag ultrathin films deposited on different substrates at condition 3 with different thicknesses of Ag.

| Film | Sheet resistance Ω/sq. | Transmittance (%) | FOM (m Ω$^{-1}$) |
|:---:|:---:|:---:|:---:|
| A6/Z2/G | 12 | 62.85 | 0.80 |
| A6/Z2/Q | 12 | 62.64 | 0.77 |
| A8/Z2/G | 5 | 59.56 | 1.12 |
| A8/Z2/Q | 5 | 60.13 | 1.23 |
| A8/Z2/PET | 15 | 47 | 0.03 |

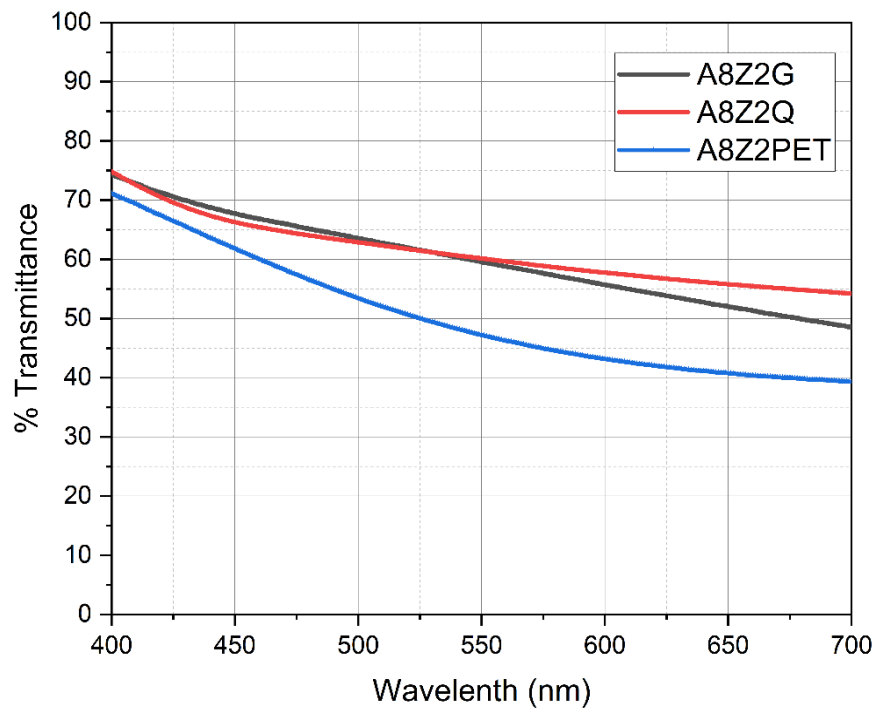

**Fig. 6.** Transmittance spectrum of 8 nm Ag film deposited on glass, quartz and PET substrate with 2 nm Zn as a filler metal at condition 3.

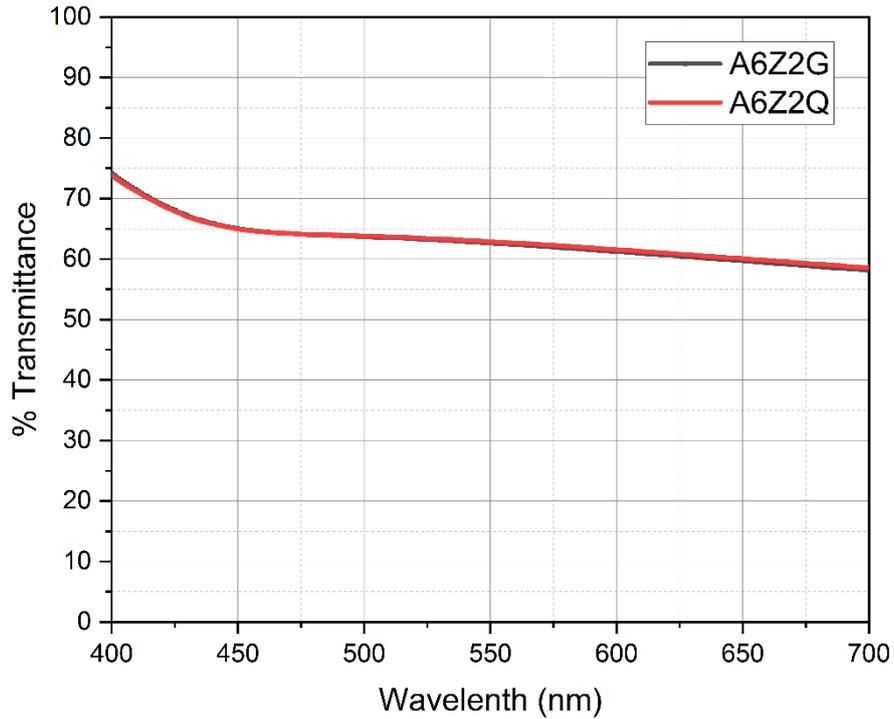

**Fig. 7.** Transmittance spectrum of 6 nm Ag film deposited on glass and quartz substrate with Zn as a filler material deposited at condition 3.

From **Fig. 6.** and **Fig. 7.** The plot shows good transmittance of ultrathin Ag films in the visible spectrum. The Ag (8 nm) films that were deposited on quartz substrate had very low sheet resistance of 5 Ohm/Sq., transmittance of 60.13 % and Hackee figure of merit (FOM) of 1.23 (mOhm$^{-1}$). The Zn assisted Ag (8 nm) films on PET substrate had sheet resistance of 15 Ohm/Sq., transmittance of 47 % and FOM of 0.03 (mOhm$^{-1}$), which is relatively low compared with the ultrathin Ag films shown in **Table 2**. These films were more stable here is the plot which shows the sheet resistance with respect to the time.

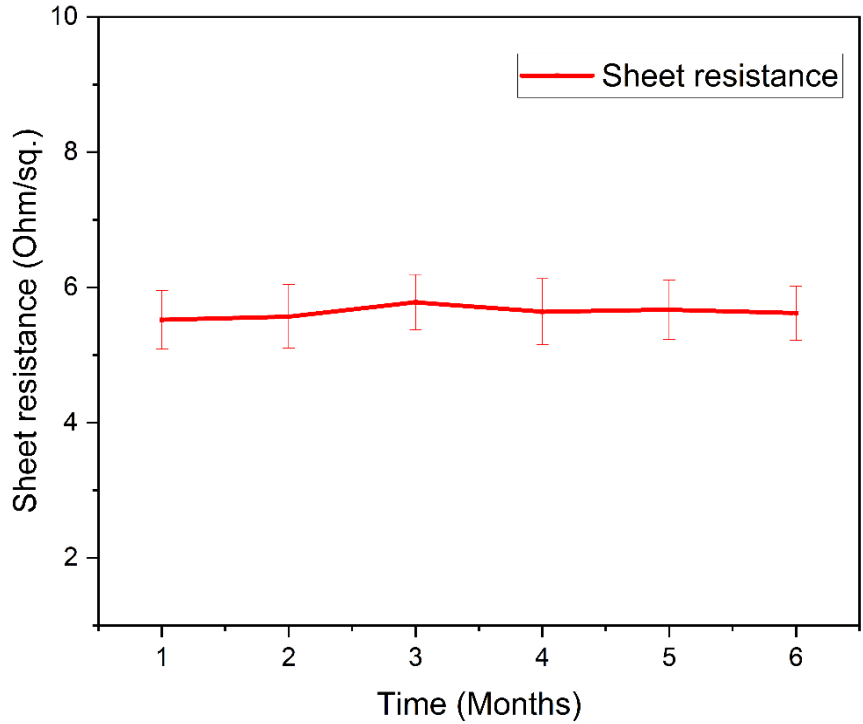

**Fig. 8.** Sheet resistance vs time of Zn (2 nm) assisted Ag (8 nm) ultrathin films deposited at condition 3.

## 4. Conclusion

From DFT results Zinc has that unique ability to absolutely get into the nano crevices more efficiently and fill them up. A Zn assisted 3D interface approach was used for successfully depositing the uniform ultrathin Ag films on glass, quartz, silicon and PET substrates. We observed that the reason for stability and good adhesion of Ag and Zn on different substrates is due to Zn filler material and ultrathin Ag film and substrate form a 3-D interface augmented atomically chemically graded interfaces. 3-D interfaces have smoothly varying chemistry. It was observed that the sheet resistance of Ag (6-8 nm)/Zn (2 nm) film on glass, quartz, silicon and PET substrates were ~ 5-15 Ω/Sq., measured mean absolute surface roughness values were between ~1-2 nm and optical transparency ~ 62 %. These films showed better stability and compatibility with the environment. Future work would focus on functionalizing the ultrathin Ag film for optoelectronics.

**Acknowledgment:**

The four-probe measurements were performed at Electronic Materials and Thin Films laboratory, at IIT Madras. Dr P Muhammed Razi helped establish the experimental set-up needed to perform DC magnetron sputtering. The AFM measurements were carried out at the CIF, ICSR, IIT Madras. The optical profilometry readings were carried out at the Centre for NEMS and Nanophotonics (CNNP),

**Author Contributions:**

**Allamula Ashok:** Conceptualization, Methodology, Investigation, Formal analysis, Writing – original draft, Writing – review and editing; **Satyesh Kumar Yadav:** Conceptualization, Supervision, Writing – review and editing.

**Conflict of Interest:**

The authors declare that they have no known competing financial interests or personal relationships that could have appeared to influence the work reported in this paper.